\documentstyle[amssymb,twocolumn,aps]{revtex}

\begin{document}
\title{Melting of the vortex lattice in high $T_{c}$ superconductors }
\author{Dingping Li$^{\text{1,2,3}}$ and Baruch Rosenstein$^{\text{1,2}}$}
\address{$^{\text{1}}${\it Electrophysics Department, National Chiao Tung}\\
University, Hsinchu 30050, Taiwan, R. O. C.}
\address{$^{\text{2}}${\it National Center for Theoretical Sciences, No.}\\
101, Sec 2, Kuang Fu Road, Hsinchu30043, Taiwan,R.O.C.\\
$^{\text{3}}${\it Department of Physics, Nanjing University,Nanjing 210093,}%
\\
China}
\date{\today}
\maketitle

\begin{abstract}
The precise measurements of vortex melting point towards a need to develop a
quantitative theoretical description of thermal fluctuations in vortex
matter. To tackle the difficult problem of melting, the description of both
the solid and the liquid phase should reach the precision level well below $%
1\%$. Such a theory in the framework of the Ginzburg - Landau approach is
presented. The melting line location is determined and magnetization,
specific heat jumps along it are calculated. We find that the magnetization
in the liquid is larger than that in the solid by $1.8\%$ regardless of the
melting temperature, while the specific heat jump is about $6\%$ and slowly
decreases with temperature. The magnetization curves agree with experimental
results on $YBCO$ and Monte Carlo simulations.
\end{abstract}




A magnetic field generates an array of vortices in type II superconductors.
The vortices strongly interact with each other forming highly correlated
configurations such as the vortex lattice. In high $T_{c}$ cuprates at
relatively high temperatures, vortices move and vibrate due to thermal
fluctuations to the extent that the lattice can melt becoming a ''vortex
liquid''\cite{Nelson}. Several recent remarkable experiments clearly
determined that the vortex lattice melting in high $T_{c}$ superconductors
is a first order phase transition with magnetization jumps\cite{Zeldov} and
spikes in specific heat \cite{Schilling}. The magnetization and entropy
jumps were measured using diverse techniques such as local Hall probes,
SQUID \cite{Pastoriza,Welp}, torque magnetometry\cite{Willemin,Nishizaki}
and integration of the specific heat spike\cite{Schilling}. Related
investigations indicate that, in addition to the spike, there is also a jump
in specific heat\cite{Schilling,Roulin}. In spite of those precise
measurements of vortex melting, a quantitative theoretical description of
vortex melting is still lacking. We present such a theory in the framework
of the Ginzburg - Landau approach in this work.

Thermal fluctuations in vortex matter have attracted great attention since
the high Tc superconductors were discovered over a decade ago. In highly
anisotropic superconductors like $BSCCO$ near the melting point vortices are
quite well separated and the system can be approximated by an array of
pointlike objects\cite{Nelson}. In less anisotropic ones like $YBCO$, near
the melting point the vortices overlap and one has to use the Ginzburg -
Landau (GL) model. The model is defined by the following free energy: 
\begin{equation}
F=\int d^{3}x\frac{{\hbar }^{2}}{2m_{ab}}\left| {\bf D}\psi \right| ^{2}+%
\frac{{\hbar }^{2}}{2m_{c}}|\partial _{z}\psi |^{2}-a(T)|\psi |^{2}+\frac{%
b^{\prime }}{2}|\psi |^{4}+\frac{\left( {\bf B}-{\bf H}\right) ^{2}}{8\pi }
\label{GLdef}
\end{equation}%
where ${\bf A}=(By,0)$ describes magnetic field (considered constant and
nonfluctuating, see below) in Landau gauge and covariant derivative is
defined by ${\bf D}\equiv {\bf \nabla }-i\frac{2\pi }{\Phi _{0}}{\bf A,}\Phi
_{0}\equiv \frac{hc}{e^{\ast }}$($e^{\ast }=2e$).When $\kappa =\lambda /\xi $
is \ large (greater than $10$ ) where $\lambda $ magnetic penetration depth
and $\xi $ coherence length, the magnetic field can be considered constant
and nonfluctuating (an excellent approximation in the region studied).
Statistical physics is described by the statistical sum \ $Z=\int D\psi D%
\overline{\psi }\exp \left( -\frac{F}{T}\right) $. It accurately describes
thermal fluctuations in the range of relatively large magnetic fields ($H>>$ 
$H_{c1}$ $=100G$ in $YBCO$) and temperatures near $T_{c}$ ($70K$ to $110K$
in $YBCO$ ). Near $T_{c}$, $a(T)$ can be approximated bt $\alpha T_{c}(1-t)$%
. In this paper, we will consider only high $\kappa $ and temperatures near $%
T_{c}$. The model is, however, highly nontrivial even within the lowest
Landau level (LLL) approximation. In this approximation, which is valid when
the magnetic field is high, only lowest Landau level mode is retained and
the free energy simplifies (after rescaling) \cite{PRL}: 
\begin{equation}
f=\frac{1}{4\pi \sqrt{2}}\int d^{3}x\left[ \frac{1}{2}|\partial _{z}\psi
|^{2}+a_{T}|\psi |^{2}+\frac{1}{2}|\psi |^{4}\right] .  \label{GLscaled}
\end{equation}%
The simplified model has only one parameter, the dimensionless scaled
temperature $a_{T}=-\left( \frac{b\omega }{4\pi \sqrt{2}}\right)
^{-2/3}a_{h} $ , where $\omega =\sqrt{2Gi}\pi ^{2}t,\ \ \ a_{h}=\frac{1-t-b}{%
2},t=T/T_{c},b=B/H_{c2}$. The dimensionless Ginzburg number $Gi$
characterizing the importance of thermal fluctuations is $\frac{1}{2}\left( 
\frac{8\pi \kappa ^{2}\xi T_{c}\gamma }{\Phi _{0}^{2}}\right) ^{2}$, and the
anisotropy parameter $\gamma $ is $\sqrt{m_{c}/m_{ab}}$.

The LLL GL model was studied by a variety of different nonperturbative
analytical methods. Among them the density functional\cite{Herbut}, $1/N$ %
\cite{largeN}, elasticity theory \cite{Maniv} and others \cite{Andreev}. The
model was also studied numerically in both 3D \cite{Sasik} and 2D \cite%
{Kato,MCother} However we will not discuss those approaches in this paper.

While applying the renormalization group (RG) on the one loop level to this
model, Brezin, Nelson and Thiaville \cite{Brezin} found no fixed points of
the (functional) RG equations and thus concluded that the transition to the
solid phase is not continuous. The RG method therefore cannot provide a
quantitative theory of the melting transition. Two perturbative approaches
were developed and greatly improved recently to describe the solid phase and
liquid phases, respectively. The perturbative approach on the liquid side
was pioneered by Ruggeri and Thouless\cite{Ruggeri}, who developed an
expansion in which all the ''bubble'' diagrams are resummed. Unfortunately,
they found that the series are asymptotic and, although first few terms
provide accurate results at very high temperatures, the series become
inapplicable for $a_{T}$ less than $-2$ which is quite far above the melting
line (located around $a_{T}=-10$). We obtained recently an optimized
gaussian series\cite{PRL} which is convergent rather than asymptotic with
radius of convergence of $a_{T}\approx -5$, but still above the melting line.

On the solid side, Eilenberger\cite{Eilenberger} calculated the fluctuations
spectrum around Abrikosov's mean field solution. Maki and Takayama\cite{Maki}
noticed that the vortex lattice phonon modes are softer than those of the
usual acoustic phonons in atomic crystals and this leads to infrared (IR)
divergences in certain quantities. This was initially interpreted as the
destruction of the vortex solid by thermal fluctuations and the perturbation
theory was abandoned. However the divergences resemble the ''spurious'' IR
divergences in the critical phenomena theory. A recent analysis demonstrated
that all these IR divergences cancel in physical quantities\cite{Rosenstein}
and the series therefore are reliable. The two loop calculation was
performed, so that the LLL GL theory on the solid side is now precise enough
even for the description of melting.

However, on the liquid side, a theory for $a_{T}\leqslant -5$ is required.
Moreover this theory should be extremely precise since the internal energies
of the solid and the liquid near melting differ by a few percents only.
Developing such a theory requires a better qualitative understanding of the
metastable phases of the model. It is clear that the overheated solid
becomes unstable at some finite temperature. It is not clear however whether
the overcooled liquid becomes unstable at some finite temperature (like
water) or exists all the way down to $T=0$ as a metastable state. Despite
its limited precision, the gaussian (Hartree - Fock) variational
calculation, is usually a very good guide to the qualitative features of the
phase diagram. While such a calculation in the liquid was performed quite
some time ago\cite{Ruggeri}, on the solid side a more complicated one
sampling inhomogeneous states was performed recently\cite{PRL}. The results
are as follows. The solid state is the stable one below the melting
temperature, becomes metastable at somewhat higher temperatures and is
destabilized at $a_{T}\approx -5$. The liquid state becomes metastable below
the melting temperature, however in contrast to the solid, it does not loose
metastability all the way down to $T=0$ and the excitation energy approaches
zero.

Meanwhile, similar qualitative results have been obtained in a different
field of physics. A variety of analytical and numerical methods\cite%
{Carvalho} have indicated that liquid (gas) phase of the classical one
component Coulomb plasma also exists as a metastable state down to $T=0$
with energy gradually approaching that of the Madelung solid and the
excitation energy diminishing. We speculate that the same phenomenon would
appear to happen in any system of particles interacting via long range
repulsive forces. In fact the vortices in the London approximation resemble
repelling particles with the force even more long range than the Coulomb. In
light of this impetus to consider the above scenario in the vortex matter,
we provide both theoretical and phenomenological evidence that the above
picture is a valid one.

Assuming the absence of singularities on the liquid branch allows to develop
a sufficiently precise theory of the LLL GL model in a vortex liquid (even
including an overcooled one) using the Borel - Pade (BP)\cite{Baker} method
at any temperature. After clarifying several issues that prevented its use
and acceptance previously we then combine it with the recently developed LLL
theory of solids\cite{Rosenstein} to calculate the melting line and the
magnetization and specific heat jumps along it. Early on, Ruggeri and
Thouless\cite{Ruggeri} tried unsuccessfully to calculate the specific heat
by using BP because their series was too short. Subsequent attempts to
calculate the melting point by using BP also ran into problems. Hikami,
Fujita and Larkin (18) attempted to find the melting point by comparing the
BP (liquid) energy with the one loop solid energy and, in doing so, obtained
the melting temperature $a_{T}^{m}=-7$. However, their one loop solid energy
was incorrect and, in any case, the two loop correction is necessary. As
demonstrated in the following, the BP energy combined with the correct two
loop solid energy computed recently not only gives $a_{T}^{m}=-9.5$, but
also allows to obtain a wide range of quantitative predictions within the
model.

Now we present the solution of the LLL GL model. The liquid LLL (scaled)
effective free energy (of the scaled model defined in eq.(\ref{GLscaled}))
is written as $f_{liq}=4\varepsilon ^{1/2}[1+g\left( x\right) ]$ . The
function $g$ can be expanded as $g\left( x\right) =\sum c_{n}x^{n}$ , where
the high temperature small parameter $x=\frac{1}{2}\varepsilon ^{-3/2}$\ is
defined as a solution of the gaussian gap equation, $\varepsilon
^{3/2}-a_{T}\varepsilon ^{1/2}-4=0$ for the excitation energy . The
coefficients can be found in ref. \cite{Hikami,Brezin1}. We will denote $%
g_{k}(x)$ by the [$k,k-1$] BP transform of $g(x)$ (other BP approximants
clearly violate the correct low temperature asymptotics). The BP transform
is defined as $\int_{0}^{\infty }g_{k}^{\prime }\left( xt\right) \exp \left(
-t\right) dt$ where $g_{k}^{\prime }$ is the $[k,k-1]$ Pade transform of $%
\sum_{n=1}^{2k-1}\frac{c_{n}x^{n}}{n!}$. For $k=4$ the liquid energy already
converges. We used in this work $k=5$ to achieve the required precision ($%
\sim 0.1\%$). The liquid energy completely agrees with the optimized
gaussian expansion results\cite{PRL} above its radius of convergence at $%
a_{T}=-5$. In addition similar results we obtained in the 2D GL model agree
with existing Monte Carlo simulations\cite{Kato}. The solid effective energy
to two loops is\cite{Rosenstein}: 
\begin{equation}
f_{sol}=-\frac{a_{T}^{2}}{2\beta _{A}}+2.848\left| a_{T}\right| ^{1/2}+\frac{%
2.4}{a_{T}}.
\end{equation}
Comparing the solid energy to that of the liquid (inset in Fig.1), reveals
that $a_{T}^{m}=-9.5$. This is in accord with experimental results. As an
example, on Fig.2 we present the fitting of the melting line of fully
oxidized $YBa_{2}Cu_{3}O_{7}$ \cite{Nishizaki} which gives $%
T_{c}=88.2,H_{c2}=175.9,Gi=7.0$ $10^{-5},\kappa =50$. Melting lines of
optimally doped untwinned \cite{Schilling} $YBa_{2}Cu_{3}O_{7-\delta }$ and $%
DyBa_{2}Cu_{3}O_{7}$ \cite{Revaz} are also fitted extremely well. For
example, for $YBa_{2}Cu_{3}O_{7-\delta }$ in \cite{Schilling}, the melting
line fitting gives $T_{c}=93.07,H_{c2}=167.53,Gi=1.9$ $10^{-4},\kappa =48.5$
(see also ref. \cite{Pierson} ).

3D Monte Carlo simulations \cite{Sasik} are not precise enough to provide an
accurate melting point since the LLL scaling is violated. One gets different
values of $a_{T}^{m}=-14.5,-13.2,-10.9$ at magnetic fields $1,2,5$T
respectively. This violation is perhaps due to a small sample size (of order
100 vortices). The situation in 2D is better since the sample size is much
larger. We performed similar calculation in the 2D LLL GL model and found
that the melting point is $a_{T}^{m}=-13.2$ . It is in good agreement with
the MC simulations\cite{Kato}. Phenomenologically the melting line can be
located using Lindemann criterion or its more refined version using Debye -
Waller factor\cite{Stevens}. The more refined criterion is required in the
case of $YBCO$ since vortices are not point-like. Numerical investigation of
the Yukawa gas\cite{Stevens} indicated that the Debye - Waller factor $%
e^{-2W}$ (a ratio of the structure function at the second Bragg peak at
melting to its value at $T=0$) is about $60\%$. We get using methods of ref. %
\cite{Licorr} $e^{-2W}=0.59$ for $a_{T}^{m}=-9.5$.

The scaled magnetization is defined by $m\left( a_{T}\right) =-\frac{d}{%
da_{T}}f_{eff}\left( a_{T}\right) $. At the melting point $a_{T}^{m}=-9.5$
the magnetization jump ratio defined by $\Delta M$ divided by the
magnetization at the melting on the solid side is found to be equal to 
\begin{equation}
\frac{\Delta M}{M_{s}}=\frac{\Delta m}{m_{s}}=.018.
\end{equation}
This prediction is compared on Fig.2 (the upper inset) with the experimental
results on fully oxidized $YBa_{2}Cu_{3}O_{7}$ (ref. \cite{Nishizaki},
rhombs) and optimally doped $YBa_{2}Cu_{3}O_{7-\delta }$ (ref. \cite{Welp},
stars) . These samples probably have the lowest amount of disorder which is
not included in the calculations. From the model, we caculate the specific
heat jump ratio at the melting 
\begin{equation}
\Delta c=0.0075\left( \frac{2-2b+t}{t}\right) ^{2}-0.20Gi^{1/3}\left(
b-1-t\right) \left( \frac{b}{t^{2}}\right) ^{2/3}.  \label{spjump}
\end{equation}
It is compared on the lower inset on Fig.2 with the experimental values of
ref. \cite{Schilling} (using the fitting parameters given above).

In addition to describing the melting, we present here an example of
quantitative results that are obtained using the present approach - the
magnetization curves. Our LLL magnetization curve coincides with the LLL
Monte Carlo result of ref. \cite{Sasik} (which is very accurate since the
LLL scaling is obeyed) to the precision of MC. However in experiments away
from the melting line higher Landau levels (HLL) are no longer negligible.
Naively in vortex solid when the distance from the mean field transition
line is smaller than the inter - Landau level gap, $1-t-b<2b$, one expects
that the higher Landau modes can be neglected. More carefully examining the
mean field solution reveals that a weaker condition $1-t-b<12b$ should be
used for a validity test of the LLL approximation \cite{LiHLL} in vortex
solids. In vortex liquid one has to go beyond the mean field to estimate the
HLL contribution \cite{Lawrie}. In 3D Lawrie in \cite{Lawrie} calculated the
excitation energy in the framework of the gaussian (Hartree - Fock)
approximation. The excitation gap is 10 times smaller than inter Landau gap
for fields in a wide range around melting line for fields larger than $0.1T$
in $YBCO$. Therefore in the range of values of the interest in the present
paper the LLL contribution should be dominant. Experimentally it is often
claimed that one can establish the LLL scaling for fields above $3$ T (see,
for example, ref. \cite{Sok}).

The theoretical expressions we use are the LLL contribution to the
magnetization plus the corrections due to HLL calculated in gaussian
approximation. The results are compared on Fig.3 with the experimental
magnetization curves of ref. \cite{Nishizaki}. We use the parameters from
the fitting of the melting curve (see Fig.2) . The agreement is fair at
higher fields, while at low magnetic fields the higher Landau levels theory
beyond gaussian approximation is required. The LLL scaling has a limited
validity away from melting line.

To summarize, the problem of the quantitative description of melting of the
vortex lattice in the framework of the LLL Ginzburg - Landau approach is
solved. The results for melting line, magnetization jump and specific heat
jump are in good agreement with experiments and MC simulations. This is the
first quantitative theory of the first order melting of any kind to our
knowledge. We believe that similar methods can be applied to other systems
undergoing the first order melting transition.

We are grateful to T. Nishizaki and A. Junod for providing details of their
experiments. The work was supported by NSC of Taiwan grant NSC$\#$%
90-2112-M-009-039.

\begin{figure}[tbp]
\caption{ Free energy of solid (line) and liquid (dashed line) as function
of the scaled temperature. The solid line ends at a point (dot) indicating
the loss of metastability. Inset shows a tiny difference between liquid and
solid near the melting point. }
\end{figure}

\begin{figure}[tbp]
\caption{ Comparison of the experimental melting line for fully oxidized $%
YBa_{2}Cu_{3}O_{7}$ in ref. \protect\cite{Nishizaki} with our fitting. Inset
on the right shows the relative universal magnetization jump of 1.8 percent
(line) and experimental results for fully oxidized $YBa_{2}Cu_{3}O_{7}$ in
ref. \protect\cite{Nishizaki} (rhombs) and optimally doped untwinned $%
YBa_{2}Cu_{3}O_{7-\protect\delta }$ in ref. \protect\cite{Welp} (stars).
Inset on the left shows the relative nonuniversal specific heat jump (line)
and experimental results for optimally doped untwinned $YBa_{2}Cu_{3}O_{7-%
\protect\delta }$ in ref. \protect\cite{Schilling}. }
\end{figure}

\begin{figure}[tbp]
\caption{ Comparison of the theoretical magnetization curves (lines) of
fully oxidized $YBa_{2}Cu_{3}O_{7}$ utilizing parameters obtained by fitting
the melting line on Fig.2 with torque magnetometry experimental results %
\protect\cite{Nishizaki} (dots). Arrows indicate melting points while at low
magnetic field the experimental data start from the point in which the
magnetization is reversible indicating low disorder. }
\end{figure}

\end{document}